\documentclass{aa} 
\usepackage{graphicx}
\usepackage{natbib}
\usepackage{txfonts}
\bibpunct{(}{)}{;}{a}{}{,}

\begin{document}

\title{Multiwavelength observations of a giant flare on CN~Leonis\thanks{Based on observations
collected at the European Southern Observatory, Paranal, Chile, 077.D-0011(A) 
and on observations obtained with \emph{XMM-Newton}, an ESA science mission 
with instruments and contributions directly funded by ESA Member States and 
NASA.}}
\subtitle{III. Temporal evolution of coronal properties}
\titlerunning{A giant flare on CN~Leo in X-rays}

\author{C. Liefke\inst{1,2}\and B. Fuhrmeister\inst{1} \and J.~H.~M.~M. Schmitt\inst{1}}

\institute{Hamburger Sternwarte, Univers\"at Hamburg, Gojenbergsweg 112, 21029 Hamburg, Germany\\ \email{cliefke@hs.uni-hamburg.de}\label{inst1} 
\and
Zentrum f\"ur Astronomie, M\"onchhofstra{\ss}e 12-14, 69120 Heidelberg, Germany \label{inst2}}

\date{Received 7 January 2010 / Accepted 13 March 2010}

\abstract
{Stellar flares affect all atmospheric layers from the photosphere over chromosphere and transition region up into the corona. Simultaneous observations in different spectral bands allow to obtain a comprehensive picture of the environmental conditions and the physical processes going on during different phases of the flare.}
{We investigate the properties of the coronal plasma during a giant flare on the active M dwarf CN~Leo observed simultaneously with the UVES spectrograph at the VLT and \emph{XMM-Newton}.}
{From the X-ray data, we analyze the temporal evolution of the coronal temperature and emission measure, and investigate variations in electron density and coronal abundances during the flare. Optical \ion{Fe}{xiii} line emission traces the cooler quiescent corona.}
{Although of rather short duration (exponential decay time $\tau_{LC} < 5$ minutes), the X-ray flux at flare peak exceeds the quiescent level by a factor of $\approx$100. The electron density averaged over the whole flare is greater than $5 \cdot 10^{11}$~cm$^{-3}$. The flare plasma shows an enhancement of iron by a factor of $\approx$2 during the rise and peak phase of the flare. We derive a size of $<9000$~km for the flaring structure from the evolution of the the emitting plasma during flare rise, peak, and decay. }
{The characteristics of the flare plasma suggest that the flare originates from a compact arcade instead of a single loop. The combined results from X-ray and optical data further confine the plasma properties and the geometry of the flaring structure in different atmospheric layers.}


\maketitle

\section{Introduction}
Flares are among the most prominent signs of stellar activity. They can be observed over the entire electromagnetic spectrum from X-ray to radio wavelengths, demonstrating that all layers of the stellar atmosphere are affected, with a wide range of plasma temperatures and densities involved.
A flare is a highly dynamic event, revealing the complex behavior of the stellar atmosphere with rapidly changing physical conditions, i.\,e. its interplay with varying magnetic fields in time and its response to the sudden release of large amounts of energy. In the commonly accepted picture of a stellar flare \citep{Haisch_Flares, Priest_Forbes}, a magnetic reconnection event in the corona drives the acceleration of particles that radiate in nonthermal hard X-rays and radio gyrosynchrotron emission as they spiral down along the magnetic field lines. Lower atmospheric layers are strongly heated by the impact of these particles and immediately start cooling by radiation in the optical and UV continuum as well as in chromospheric and transition region emission lines. Chromospheric evaporation then brings ``fresh material'', emitting soft thermal X-ray emission, into the corona. The strong X-ray and UV radiation field finally induces further chromospheric emission \citep{Hawley_Fisher}. 

Flares on the Sun and on stars show a wide variety of amplitudes, from the smallest micro/nanoflares to giant flares with luminosity increases by orders of magnitude, and timescales ranging from a few seconds up to several days. Depending on energy budget, decay time and shape of the lightcurve, impulsive and gradual flares can be discerned. On the Sun, the former have been associated with compact emission regions, i.\,e., single loops, while the latter typically involve a series of eruptions in a whole arcade of loops \citep[so-called two-ribbon flares, see][]{Pallavicini_flaretypes}. For stellar flares, the loop geometry can usually not be observed directly, but scaling laws based on the hydrostatic case, relating the loop temperature and pressure with the size of the loop, as derived and tested for the quiet Sun \citep{RTV} can be adapted to stellar flares to give an estimate of the dimensions of the involved coronal structures \citep{Aschwanden_laws}. Additionally, hydrodynamic loop modeling approaches allow to assess the loop length from the analysis of the decay of the flare in X-rays and to estimate the amount of additional heating \citep{Serio_flaremodel, Reale_loops}. This approach has recently been extended to an analysis of the rise phase of the flare \citep{Reale_diagnostics}. Individual flare events can thus be characterized with respect to to the physical properties of the flaring structure from observational quantities also available for stellar events. The greater the spectral coverage and the better the spectral and temporal resolution, the more information can be gained and the better are the constraints on the physical properties of the flare plasma and the geometry of the flaring active region. Simultaneous observations in multiwavelength bands allow to study the flare throughout all atmospheric layers from the photosphere into the corona.

Strong flares dominate the overall flux level in all wavelength bands; the X-ray luminosity can increase by more than two orders of magnitude during the strongest events \citep[e.\,g.][]{Algol_nature, Algol_Flare, Favata_EV_Lac}, which allows the strongest events even on relatively faint targets to be investigated in great detail. For outstanding flares like the long-lasting giant flare observed on Proxima~Cen with \emph{XMM-Newton}, the temporal evolution of plasma temperatures, densities and abundances during different flare phases could be observed directly \citep{Guedel_Proxima_Cen_1, Guedel_Proxima_Cen_2}; detailed and specific hydrodynamic loop modeling revealed a complex flare geometry with two loop populations and differing heating mechanisms \citep{Reale_Proxima_Cen}.

Even for much smaller flares, low- and intermediate-resolution X-ray spectroscopy makes it possible to follow the evolution of plasma temperature and emission measure during flare rise and decay. The temperature of the flare plasma is typically observed to peak before the maximum emission measure is reached, verifying the overall framework of initial heating, evaporation, and conductive and radiative cooling as predicted by hydrodynamic loop models \citep[e.\,g.][]{Reale_diagnostics}. High-resolution spectroscopy is needed to directly measure coronal densities during flares and thus to obtain the weights of changes in plasma density or volume. In general, coronal densities deduced from the ratio of the forbidden and intercombination lines in He-like triplets, especially from \ion{O}{vii} and \ion{Ne}{ix} at formation temperatures of 2~MK and 3.5~MK, respectively, during smaller flares are consistent with the quiescent value within the errors but tend to be higher \citep[see e.\,g.][]{Mitra_Kraev_densities}; unambiguously increased coronal densities have so far only been measured during the giant flare on Proxima Centauri \citep{Guedel_Proxima_Cen_1}. 

Another aspect of the evaporation scenario is the observation of abundance changes during flares, leading directly to the question to what extent flares influence the overall coronal abundance pattern. \citet{Nordon_flares, Nordon_flares2} systematically analyzed abundance trends in a sample of flares observed with \emph{Chandra} and \emph{XMM-Newton} and found that the most active stars with a pronounced inverse FIP effect during quiescence compared to solar photospheric values showing a shift to a FIP effect compared to the corresponding quiescent values, and vice versa for the less active stars showing the solar-like FIP effect during quiescence. This confirms the existence of the inverse FIP effect and of strong discontinuities in the abundance patterns of corona and chromosphere/photosphere also for stars where photospheric abundances are difficult to measure, however, the question what causes this bias is left unanswered .

We present the observation of a giant flare on the active M~dwarf CN~Leo with \emph{XMM-Newton} and the UVES spectrograph at the VLT. An extremely short impulsive outburst at the beginning of the flare has already been discussed by \citet{CN_Leo_onset}. This paper is the concluding third paper of a series analyzing the whole flare event in greater detail. Paper~I \citep{flare_chromos} reports the observational results on the chromospheric emission as diagnosed by the UVES spectra, while Paper~II \citep{flare_model} describe model chromospheres and synthetic spectra obtained with the atmospheric code \emph{PHOENIX}. Here we focus on the temporal evolution of the coronal plasma temperature, emission measure and abundances determined by the X-ray data and their implications on the geometry of the flaring structure, on the physical conditions dominating the flare environment, and on the abundance patterns predominating in different atmospheric layers.

\section{The active M dwarf CN Leo}
\object{CN Leo} (GJ~406, Wolf~359) is, after the $\alpha$~Centauri system and Barnard's star, the fifth-nearest star to the Sun at a distance of 2.39~pc. Despite its proximity, it is a faint object with a V band magnitude of 13.54; recent spectral classification ranks it as an intermediate M~dwarf of spectral type M5.5 \citep{Reid_Mdwarfsurvey} or M6.0 \citep{Kirkpatrick_spectraltypes}. \citet{Fuhrmeister_Mdwarfmodels} fitted $T_\mathrm{eff}$=2900\,K and $\log g = 5.5$ based on a model grid with $\Delta T_\mathrm{eff} = 100$~K and $\Delta\log g = 0.5$, while \citet{Pavlenko_CN_Leo} found $T_\mathrm{eff} = 2800 \pm 100$~K at a fixed $\log g$ of 5.0 and solar metallicity. From  $M_\mathrm{bol} = 12.13 \pm 0.10$ the latter determined a mass of 0.07--0.1~$M_\odot$ and a rather young age of 0.1--0.35~Gyr, consistent the lower limit of 0.1~Gyr deduced from the absence of lithium signatures \citep{Magazzu_Lithium}. However, based on its kinematics, \citet{Delfosse_mdwarfs} and \citet{Mohanty_Basri} classify CN~Leo as intermediate young/old disk or old disk star.

Nevertheless, CN~Leo is well-known to show several signs of activity. \citet{Lacy_flarestars} and \citet{Gershberg_flarestars} found high optical flare rates, however, at the lowest flare energies in their samples of UV~Ceti stars. \citet{Robinson_CN_Leo} analyzed UV flare events on CN~Leo down to the microflare scale observed with the High Speed Photometer onboard \emph{HST}.
From 6~cm and 20~cm VLA observations, \citet{ODea_CN_Leo_radio} could only specify a $3\sigma$ upper limit on its radio flux, thus CN~Leo is underluminous in this wavelength regime compared to other active M~dwarfs. 
\citet{Guedel_correlation} were able to detect CN~Leo in the 3.5~cm and 6~cm bands during a flare in a simultaneous observing campaign with the VLA at radio wavelengths and with the \emph{ROSAT} PSPC in X-rays. During quiescence, however, again only upper limits were found at radio wavelengths. \citet{Reale_loops2} discussed the X-ray data of the flare and derived a loop half-length of $\approx 7 \cdot 10^9$~cm from the analysis of the flare decay. 
The first X-ray detections of CN~Leo have already been reported from \emph{Einstein} data \citep{Vaiana, Golub_Einstein_stars}, and even during quiescence, it is a relatively strong X-ray source with $\log L_\mathrm{X}=26.97$ and 27.01 in the \emph{ROSAT} All-Sky survey and a second pointed PSPC observation, respectively \citep{NEXXUS}. CN~Leo is the only star where persistent but variable optical coronal \ion{Fe}{xiii} emission has been confirmed \citep{Schmitt_FeXIII, Fuhrmeister_FeXIII}. \ion{Fe}{xiii} is an indicator for cool coronal plasma ($\approx 1.5$~MK),
while typical coronal emission measure distributions of earlier M~dwarfs peak around 7--8~MK \citep[e.\,g.][]{Robrade_M_dwarfs}, i.\,e., temperatures where the ionization equilibrium is shifted towards \ion{Fe}{xvii}. \citet{Fuhrmeister_CN_Leo} found significant amounts of cooler plasma in the corona of CN~Leo from X-ray spectra obtained with \emph{XMM-Newton}, consistent with the \ion{Fe}{xiii} line fluxes from simultaneous optical spectroscopy.
In addition to its high X-ray luminosity, CN~Leo shows strong H$\alpha$ emission \citep[$\log L_{\mathrm{H}\alpha}/L_{\mathrm{bol}} = -3.89$ or $-3.39$:][]{Mohanty_Basri, Reiners_Basri_stars} and a multitude of other chromospheric emission lines \citep{Fuhrmeister_FeXIII}. \citet{Reiners_Basri_stars} measured an integrated magnetic surface flux of 2.4~kG, that varies by $\approx 100$~G on timescales ranging from hours to days \citep{Reiners_CN_Leo}.

\section{Observations and data analysis}
The giant flare on CN~Leo discussed in this series of papers has been observed during a multiwavelength campaign on this star in X-rays and in the optical. This campaign consisted of six half nights of optical high-resolution spectroscopy monitoring with UVES at the VLT, and simultaneous \emph{XMM-Newton} observations. The first three observations have been performed on 19/20 May 2004, and 12 and 13 December 2005; they are discussed in detail by \citet{Fuhrmeister_CN_Leo}. The last set of observations has been performed on 19/20, 21/22, and 23/24 May 2006; here we focus on the giant flare that occurred in the first of these three nights. 

\emph{XMM-Newton} consists of three co-aligned X-ray telescopes, accompanied by the Optical Monitor OM, an optical/UV telescope that can be used with different filters in imaging or fast readout mode. The OM data of the flare, obtained in fast mode with the U band filter, has already been discussed by \citet{flare_chromos} in conjunction with the behavior of the optical chromospheric emission lines.
\emph{XMM}'s X-ray telescopes are equipped with EPIC (European Photon Imaging Camera) detectors, X-ray CCDs providing medium-resolution imaging spectroscopy with $E/\Delta E \approx 20$--50 and timing analysis with a time resolution at the subsecond level. There are two identical EPIC MOS detectors (MOS~1 and MOS~2), operating in the energy range of 0.2--12.0~keV, and one EPIC PN detector, which covers the energy range of 0.2--15.0~keV. The PN has a higher sensitivity, while the MOS detectors provide better angular and spectral resolution; several filters and operating modes are available for these instruments. The two X-ray telescopes with the MOS detectors are additionally equipped with reflective gratings and their corresponding CCD detectors. The Reflection Grating Spectrometers (RGS1 and RGS2) provide high-resolution spectroscopy in the energy range of 0.35--2.5~keV (5--38~\AA) with $E/\Delta E$ ranging between 200 and 800 and a spectral resolution of $\approx$0.06~\AA\ FWHM which allows to resolve individual emission lines. All instruments are usually operated simultaneously.\footnote{Further details on the 
instruments onboard \emph{XMM-Newton} can be found in the \emph{XMM-Newton} 
Users' Handbook, available at http://xmm.vilspa.esa.es/external/xmm$\_$user$\_$support/documentation/ uhb/index.html}

In the \emph{XMM-Newton} observation of 19/20 May 2006 (ObsID 0200530501), the EPIC detectors were operated in Full Frame (MOS) and Large Window mode (PN) with the medium filter. All X-ray data were reduced with the \emph{XMM-Newton} Science Analysis System (SAS) software, version 7.0. Spectral analysis of the EPIC data was carried out with XSPEC V12.3 \citep{xspec12}, making use of multi-temperature component models with coupled abundances for each component. The plasma models assume a collisionally-ionized, low-density optically-thin plasma as calculated with the APEC code \citep{APEC, APED}. Abundances are calculated relative to solar photospheric values from \citet{Anders_Grevesse}. 
Individual line fluxes in the RGS spectra have been measured with the CORA program \citep{CORA} using Lorentzian line profiles with a fixed line width of 0.06~\AA.

UVES (Ultra-violet and Visible Echelle Spectrograph) is one of the high-dispersion spectrographs of the VLT, operating from 3000 to 11\,000~\AA\ with a maximum spectral resolution of 110\,000\footnote{A detailed description of the UVES spectrograph is available under  http://www.eso.org/instruments/uves/doc/}. It was used in a non-standard setup using dichroic beam splitter No.~2 (DIC2), with spectral coverage from 3050~\AA\ to 3860~\AA\ in the blue arm and from 6400~\AA\ to 8190~\AA\ and 8400~\AA\ to 10080~\AA\ in the red arm at a spectral resolution of $\approx$40\,000. Exposure times were 1000~s in the blue arm and 200~s in the red arm. Here we discuss only the temporal behavior of the optical coronal \ion{Fe}{xiii} line at 3388~\AA\ during the flare, which is covered by the blue arm spectra. The UVES spectra were reduced using the IDL-based {\tt REDUCE} package \citep{REDUCE}. The wavelength calibration was carried out using Thorium-Argon spectra and resulted in an accuracy of $\approx$0.03~\AA\, in the blue arm and $\approx$0.05~\AA\, in the red arm. Absolute flux calibration was carried out using the UVES master response curves and extinction files provided by ESO. The flux in the \ion{Fe}{xiii} line has been measured with CORA.

\section{The Flare}
\subsection{Timing analysis}
\begin{figure}
\resizebox{\hsize}{!}{\includegraphics{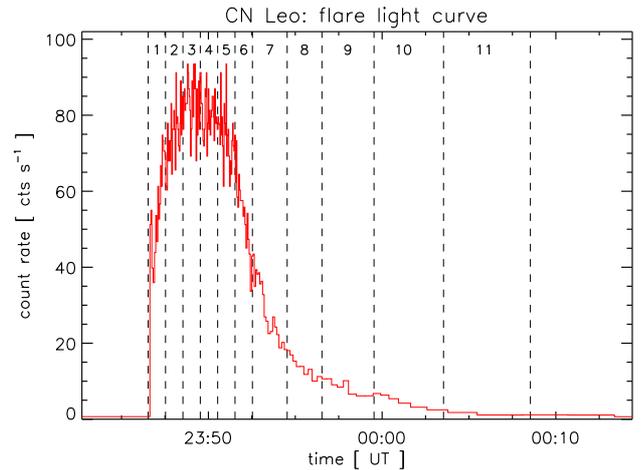}}
\caption[]{\label{lc}EPIC PN lightcurve of large flare, corrected for pileup as described in the text. Time bin sizes are chosen to contain 100 counts per bin before scaling the annular region to the total level. Time intervals used for spectroscopic analysis are marked.}
\end{figure}

The \emph{XMM-Newton} observation of CN~Leo from 19/20 May 2006 covers the giant flare discussed here, and two additional larger flares. The quasi-quiescent X-ray emission surrounding the three flares sums up to 20~ks, while the giant flare lasted $\approx$1.4~ks. The countrate increased from quiescent values of $\approx$0.7~cts/s in the PN and $\approx$0.14~cts/s in the two MOS detectors to $\approx$65~cts/s and $\approx$14~cts/s, respectively, at the flare peak. Due to the long frame times of the EPIC imaging modes used, not only the spectral response but also the shape of the lightcurve is heavily affected by pileup. From the ratio of countrates during flare and quiescence from circular and annular shaped extraction regions (the latter excluding the innermost part of the point spread function PSF), we determined that the amount of pileup reached values up to 30\% for the PN at the time of maximum countrate. The effect on the two MOS detectors is even stronger.
Figure~\ref{lc} shows the EPIC PN lightcurve extracted from an annular extraction region with the innermost 150 pixels (corresponding to the innermost 7.5$''$ of the PSF) excluded. The lightcurve has been scaled by the ratio of the quiescent countrates in circular and annular extraction regions in order to visualize the true countrates.

The initial rise of the few-second outburst at the beginning of the flare at 23:46:40~UT discussed by \citet{CN_Leo_onset} was very steep, after its short decay, the main flare continued to rise less steeply. However, still within less than two minutes, the maximum countrate was reached. Even with the effects of pileup eliminated, the flare peak is flattened, resulting in a kind of plateau of about 3 minutes duration. The subsequent decay phase can be fitted with an exponential decay time $\tau_{LC}$ of 257$\pm$13~seconds, however, it shows slight deviations from an exponential shape. Especially, during the first two minutes after the plateau phase, the decay seems to have been faster, and an additional bump peaked at 0:00~UT. Less than 25 minutes after the initial outburst, CN~Leo has returned to the quiescent level before the flare event. Stellar X-ray flares of that amplitude typically last much longer, i.\,e. from several hours up to days \citep[see e.\,g.][]{Kuerster_CF_Tuc, Algol_nature, Favata_EV_Lac, Guedel_Proxima_Cen_2}. Integrating the total energy flux in the 0.2--10.0~keV band, yields a radiative loss in the X-ray regime of $E_X \approx 4.4 \cdot 10^{31}$~erg, which can be considered to represent the major contribution of the total radiative loss. Though very energetic, the short dynamical time scales already suggest a very compact structure. The large flare is followed by a much smaller one at 0:10~UT (X-ray amplitude $\approx$2, duration $\approx$3 minutes), that is also visible in the optical (see Figure 1 in Paper~I).

\begin{figure}
\resizebox{\hsize}{!}{\includegraphics{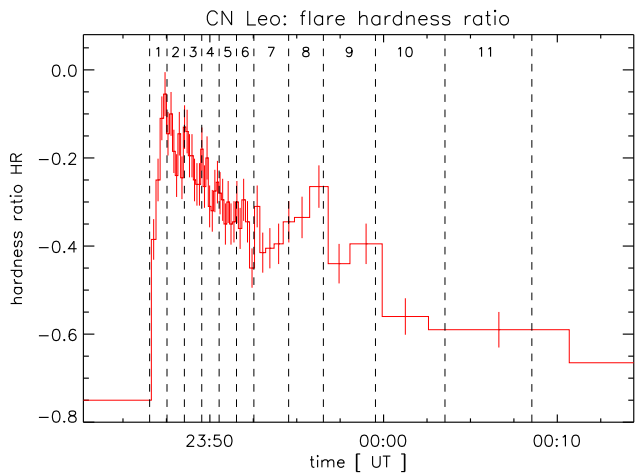}}
\caption[]{\label{hr}Temporal evolution of the hardness ratio. Time bin sizes are chosen to contain 400 counts per bin in the total $= H+S$ band.}
\end{figure}

In Fig.~\ref{hr} we show the temporal evolution of the spectral hardness during the flare. We define the hardness ratio $HR$ as $HR = \frac{H-S}{H+S}$, with the hard and soft bands ranging from 
1.0--15.0~keV and 0.15--1.0~keV, respectively. To avoid pileup effects, the annular extraction region has been used for both spectral bands. The shape of the HR lightcurve is fairly different from the normal lightcurve. Before the flare, the hardness ratio is $\approx$$-0.75$, well consistent with what \citet{Fuhrmeister_CN_Leo} found for the low-activity state of CN~Leo on 19 May 2004. With the flare rise, the hardness ratio steeply increases, peaking at a value of $\approx$$-0.05$ about one minute before the plateau phase of the lightcurve is reached. At the maximum of the HR lightcurve, the contribution from X-ray photons above the dividing line of 1~keV therefore almost equalled the amount of photons below, while during quiescence, the spectral energy distribution is by far dominated by lower energies. \citet{Guedel_Proxima_Cen_2} found a very similar behavior for the large flare on Proxima~Cen. The decay of the hardness ratio is interrupted by a new rise, peaking at 23:56~UT, approximately the time when the small bump in the lightcurve starts to emerge.

\subsection{Plasma temperature and emission measure}
\label{spectra}
\begin{figure}
\resizebox{\hsize}{!}{\includegraphics{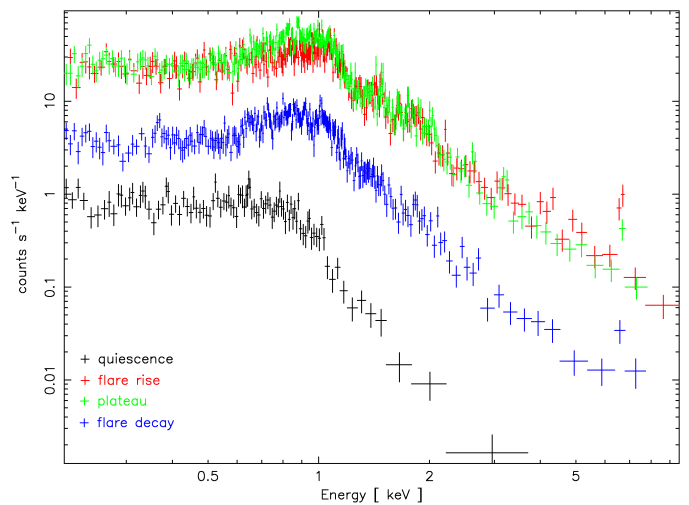}}
\caption[]{\label{EPIC}EPIC PN spectra of the quiescent state before the flare (black), of the rise phase (red, time intervals 1 and 2), of the flare peak (green, time intervals 3 and 4), and of the decay phase (blue, time intervals 5 to 11).}
\end{figure}

The striking changes of the spectral energy distribution are clearly visible in the X-ray spectra. Figure~\ref{EPIC} compares the EPIC PN spectra of the quiescent state before the flare, of the flare rise, plateau phase and decay. The quiescent spectrum is very soft, even somewhat softer than the spectrum from 19 May 2004. \citet{Fuhrmeister_CN_Leo} found CN~Leo to show three different quasi-quiescent X-ray flux levels in May 2004 and on 11 and 13 December 2006, and assumed that the two enhanced states in December 2006 reflect the decay phases of one or two long-duration flares. Our new observations now confirm the low-level quiescent state, which is also consistent with the previous \emph{ROSAT} and \emph{Einstein} data.

We fitted the quiescent spectra before and after the flare with two-temperature component APEC models. While the temperatures and emission measures proved to be stable, rather large uncertainties for the individual elemental abundances occurred with the abundances treated as free parameters. We therefore chose to adopt the values of \citet{Fuhrmeister_CN_Leo}, i.\,e. C/H~=~1.49, N/H~=~1.28, O/H~=~0.62, Ne/H~=~1.20, Mg/H~=~0.85, Si/H~=~0.78, S/H~=~1.41, and Fe/H~=~0.59 relative to the solar photospheric values of \citet{Anders_Grevesse}. The properties of the two quiescent spectra are very similar, however, for the second quiescent phase the fit quality can be improved with a three-temperature component model. In Table~\ref{quiescence} we summarize the properties of the two spectra and their associated fit parameters.

\begin{table}
\begin{center}
\caption{\label{quiescence}Fit parameters to the quiescent spectra before and after the flare}
\begin{tabular}{lccc}
\hline\hline
& before the flare & \multicolumn{2}{c}{after the flare}\\
\hline
Exposure time [s]			&3004			&\multicolumn{2}{c}{5046}\\
$kT_1$ [keV]				&0.14$\pm$0.01		&0.19$\pm$0.01	&0.13$\pm$0.01\\
$EM_1$ [10$^{50}$ cm$^{-3}$]		&0.16$^{+0.03}_{-0.02}$	&0.21$\pm$0.01	&0.16$\pm$0.02\\
$kT_2$ [keV]				&0.40$^{+0.09}_{-0.02}$	&0.58$\pm$0.02	&0.34$\pm$0.01\\
$EM_2$ [10$^{50}$ cm$^{-3}$]		&0.21$^{+0.02}_{-0.04}$	&0.13$\pm$0.01	&0.22$^{+0.01}_{-0.02}$\\
$kT_3$ [keV]				&--			&--		&1.11$^{+0.19}_{-0.18}$	\\
$EM_3$ [10$^{50}$ cm$^{-3}$]		&--			&--		&0.04$\pm$0.01\\
red. $\chi^2$				&1.05			&1.40		&1.05\\
d.o.f.					&96			&149		&147\\
$\log L_\mathrm{X}$ [0.2--5.0~keV]	&26.78			&26.77		&26.80\\
\hline
\end{tabular}
\end{center}
\end{table}

\citet{CN_Leo_onset} already showed that the flare plasma of the initial short-duration outburst is thermal and consistent with that of the following rise phase of the main flare. The large increase in temperature and emission measure during the flare rise compared to the quiescent state is obvious from Fig.~\ref{EPIC}; the iron K complex clearly emerges. Enhanced line emission between 0.6 and 1.1~keV sets in, when the plateau phase of the flare peak is reached, while during the decay both emission measure and temperatures slowly decrease. 

\begin{figure*}
\resizebox{\hsize}{!}{\includegraphics{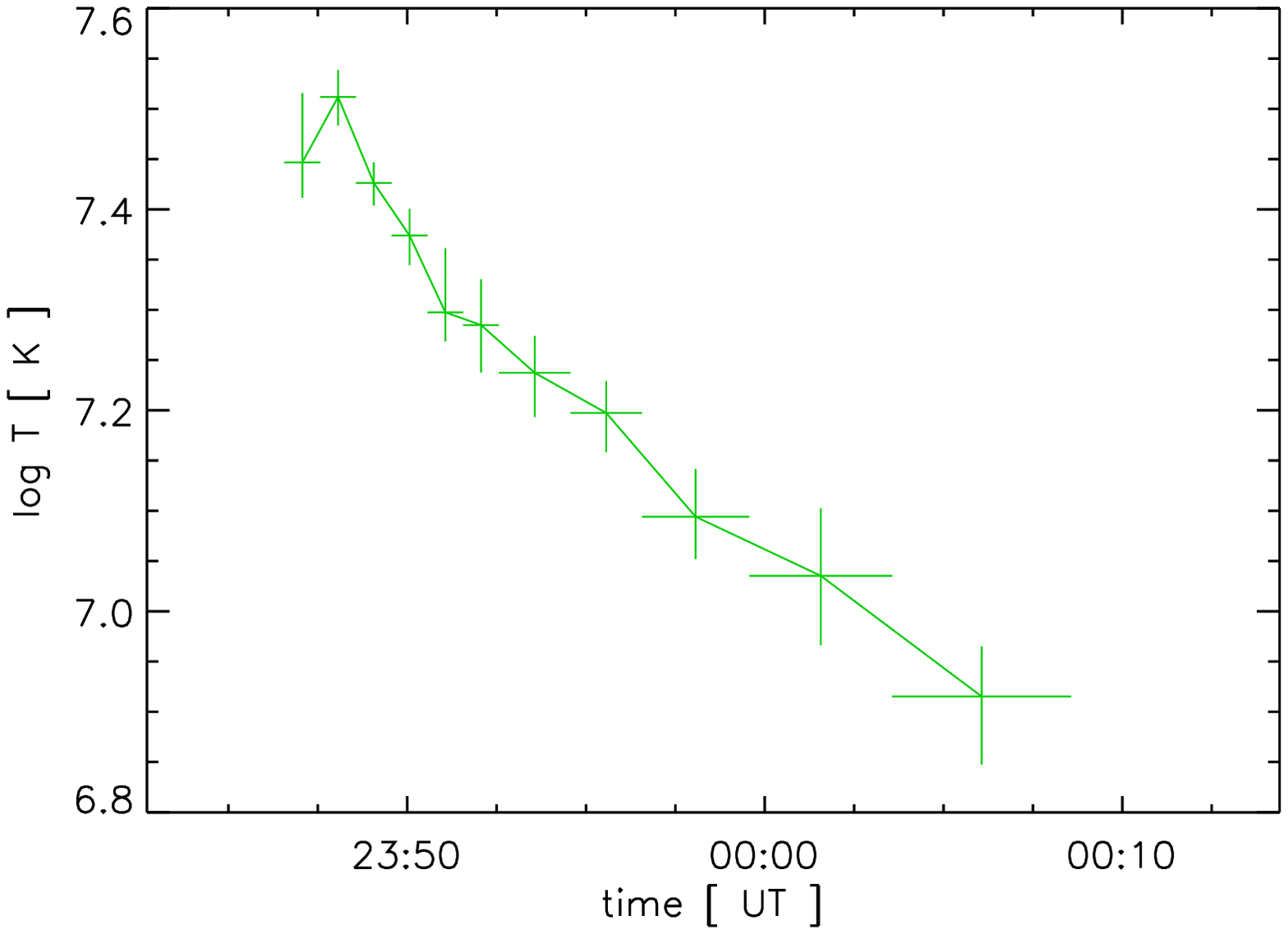}\includegraphics{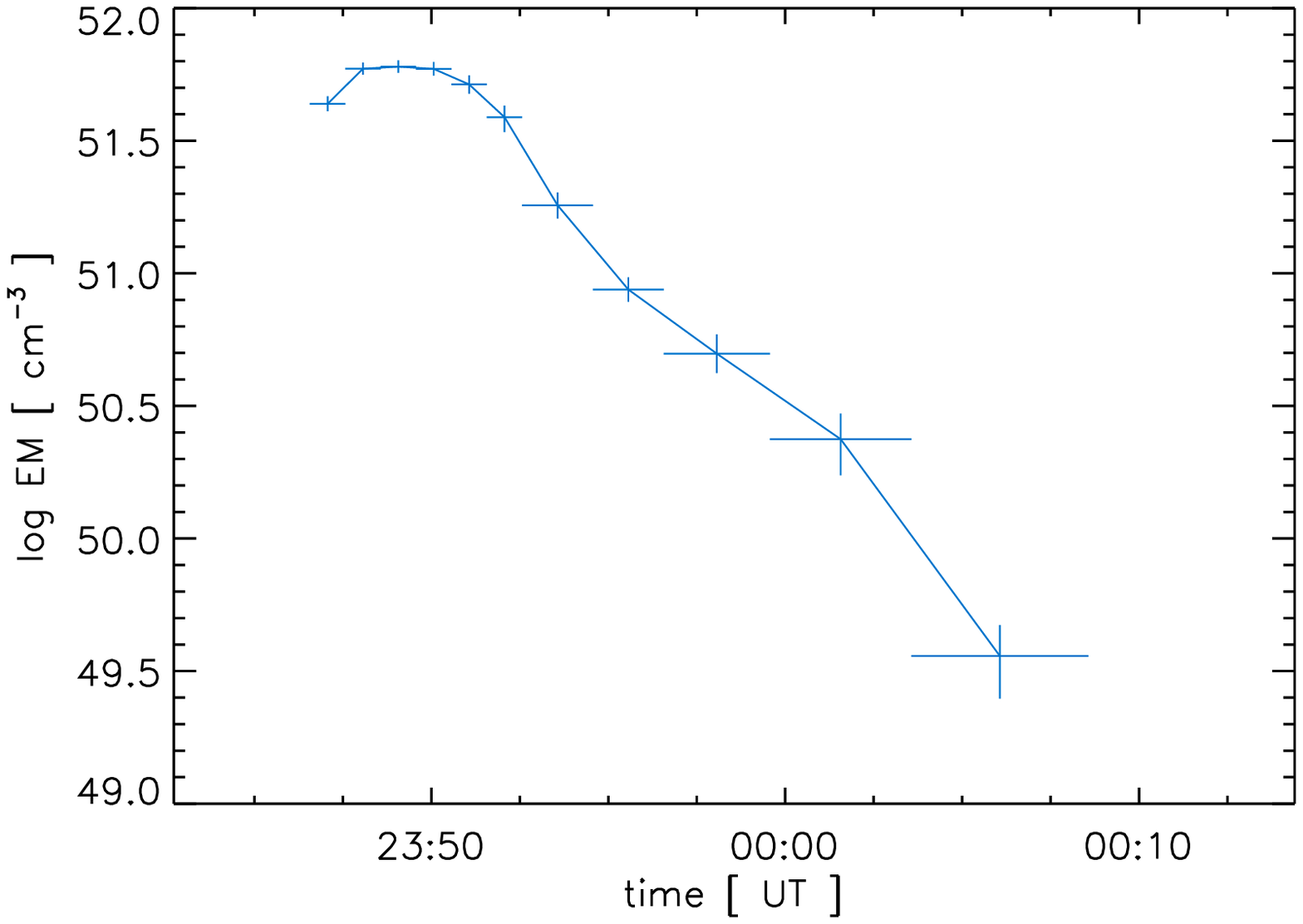}}
\caption[]{\label{EPIC_time}Temporal evolution of flare temperature (left) and total emission measure (right) of the flare plasma obtained from a two-temperature component fit with variable Fe abundance.}
\end{figure*}

\begin{table*}
\begin{center}
\caption{\label{epic_fit}Model parameters for the flare spectra with variable Fe abundance}
\begin{tabular}{cccccccccccc}
\hline\hline
Spect.& Exp. & $kT_3$&$EM_3$ &$kT_4$ &$EM_4$&Fe &$\log L_\mathrm{X}$&red. $\chi^2$&d.o.f.&$T$&$EM$\\
No.	& time [s] 		&[keV]	&[10$^{50}$ cm$^{-3}$]&[keV]	&[10$^{50}$ cm$^{-3}$]&& [0.2--10.0~keV]&&&[MK]&[10$^{50}$ cm$^{-3}$]\\
\hline
1	&60	&0.74$_{-0.05}^{+0.04}$	& 4.81$_{-0.92}^{+1.18}$&2.62$_{-0.21}^{+0.46}$&38.76$_{-1.74}^{+1.76}$&1.03$_{-0.19}^{+0.32}$	&28.90	&0.99	&91     &27.97$_{-4.82}^{+2.18}$&43.58$_{-2.66}^{+2.94}$    \\
2	&60	&0.84$_{-0.02}^{+0.09}$	& 6.99$_{-1.02}^{+1.26}$&3.06$_{-0.20}^{+0.19}$&52.11$_{-1.95}^{+1.98}$&1.38$_{-0.20}^{+0.21}$	&29.07	&1.22	&130    &32.50$_{-2.07}^{+2.06}$&59.10$_{-2.97}^{+3.23}$      \\
3	&60	&0.75$\pm$0.02		& 9.57$_{-1.13}^{+1.32}$&2.59$_{-0.13}^{+0.13}$&50.60$_{-1.98}^{+2.01}$&1.33$_{-0.15}^{+0.17}$	&29.08	&1.23	&141    &26.69$_{-1.29}^{+1.34}$&60.18$_{-3.12}^{+3.33}$    \\
4	&60	&0.73$\pm$0.03		&10.97$_{-1.38}^{+1.60}$&2.34$_{-0.16}^{+0.15}$&48.06$_{-2.03}^{+2.04}$&1.07$_{-0.13}^{+0.14}$	&29.05	&1.09	&132    &23.66$_{-1.49}^{+1.55}$&59.03$_{-3.40}^{+3.64}$      \\
5	&60	&0.75$_{-0.03}^{+0.02}$	&14.46$_{-1.85}^{+2.15}$&2.08$_{-0.14}^{+0.37}$&37.13$_{-2.11}^{+2.06}$&0.89$_{-0.11}^{+0.09}$	&28.99	&1.10	&121    &19.84$_{-3.13}^{+1.27}$&51.59$_{-3.97}^{+4.21}$     \\
6	&60	&0.71$_{-0.07}^{+0.03}$	&10.67$_{-2.68}^{+2.19}$&2.02$_{-0.21}^{+0.24}$&28.15$_{-2.01}^{+1.89}$&0.72$_{-0.12}^{+0.14}$	&28.84	&1.02	&91     &19.27$_{-2.13}^{+1.98}$&38.82$_{-4.69}^{+4.09}$   \\
7	&120	&0.71$_{-0.05}^{+0.03}$	& 6.32$_{-0.97}^{+1.16}$&1.91$_{-0.20}^{+0.18}$&11.75$_{-1.01}^{+0.95}$&0.75$_{-0.11}^{+0.13}$	&28.52	&1.26	&91    &17.27$_{-1.52}^{+1.66}$&18.07$_{-1.98}^{+2.10}$   \\
8	&120	&0.61$_{-0.07}^{+0.06}$	& 1.64$_{-0.31}^{+0.39}$&1.53$_{-0.13}^{+0.11}$& 7.05$_{-0.57}^{+0.58}$&0.85$_{-0.15}^{+0.18}$	&28.20	&0.89	&46    &15.75$_{-1.19}^{+1.35}$& 8.69$_{-0.87}^{+0.97}$   \\
9	&180	&0.58$_{-0.08}^{+0.07}$	& 1.64$_{-0.35}^{+0.49}$&1.31$_{-0.11}^{+0.15}$& 3.33$_{-0.42}^{+0.41}$&0.59$_{-0.12}^{+0.14}$	&27.94	&0.91	&40    &12.42$_{-1.43}^{+1.14}$& 4.98$_{-0.77}^{+0.90}$   \\
10	&240	&0.55$_{-0.10}^{+0.06}$	& 1.24$_{-0.31}^{+0.35}$&1.35$_{-0.18}^{+0.27}$& 1.13$_{-0.32}^{+0.24}$&0.52$_{-0.13}^{+0.24}$	&27.62	&0.95	&28   &10.85$_{-1.81}^{+1.59}$& 2.37$_{-0.64}^{+0.59}$    \\
11	&300	&0.71$_{-0.10}^{+0.09}$	& 0.36$\pm$0.11 	&--			&--			&0.49$_{-0.20}^{+0.36}$	&26.84	&0.80	&10   & 8.23$_{-1.00}^{+1.19}$& 0.36$\pm$0.11\\
\hline
\end{tabular}
\end{center}
\end{table*}

We divided the flare into eleven time intervals, the first six lasting 60~s, followed by two intervals of 120~s, and three intervals of 180~s, 240~s, and 300~s each. The first two intervals cover the flare rise, the next two the flattened peak, and the following seven intervals the different phases of the decay until the small flare at 0:10~ UT sets in. We created spectra for each time interval and fitted them with various combinations of APEC models, using different numbers of temperature components and sets of variable elemental abundances as described in the following. Our models always include the quiescent emission, i.\,e. the parameters of the first two temperature components are fixed to the plasma properties of the quiescent spectrum before the flare as listed in Table~\ref{quiescence}. With this approach, we account for the contribution of the quiescent corona to the overall X-ray emission and neglect only the quiescent emission of the active region (flaring loop or arcade), where the flare takes place. This affects primarily the analysis of the last phases of the decay where the plasma properties have almost returned to the quiescent state, while the contribution of the quiescent emission is negligible from the flare rise to the middle of the decay.

The eleven flare spectra have then been fitted with one, two, and three additional temperature components. The last spectrum however contains very few counts, considering the resulting few degrees of freedom, only a one-temperature component fit is appropriate. 
In order to compare the plasma properties of the different sets of models with $n$ additional temperature components, we make use of the total the emission measure $EM$, i.\,e. the sum of the emission measures of each temperature component $i$
\begin{equation}
EM = \sum_{i=3}^{n+2} EM_i
\end{equation}
and the flare temperature $T$,
\begin{equation}
T = \sum_{i=3}^{n+2} \frac{T_i \cdot EM_i}{EM}
\end{equation}
defined as an emission measure weighted sum of the temperatures from each flare component. 

In a first approach, we fixed the elemental abundances for the flare plasma at the values listed above. One additional temperature component turns out to fit the last two spectra at the end of the decay well, while rather large residuals remain until the middle of the decay. However, already two temperature components give a reasonable fit for these phases of the flare. Even for the well-exposed rise and peak spectra, the addition of a third component does not provide an improvement in terms of $\chi^2$, and models with two and three temperature components provide consistent flare temperatures and total emission measures.

The flare temperature peaks sharply at $\approx$35~MK in the second spectrum, and immediately starts to decay. For the first six spectra, the flare-temperature lightcurve resembles the hardness-ratio lightcurve very well, which reflects the high-temperature continuum enhancement at higher energies. In the later phases of the flare, line emission becomes more important, disturbing the direct relationship between hardness ratio and temperature. 
The total emission measure shows the flat top already known from the lightcurve, and also the bump at 0:00~UT. Compared to the flare temperature, the decay is delayed by about two minutes. 

When using two additional temperature components to fit the flare plasma, $T_3$ varies between 6 and 11~MK, while the hotter component $T_4$ covers a temperature range between 15 and 45~MK, with both temperature components decreasing starting with the second time interval. The ratio of the two corresponding emission measures $EM_3/EM_4$, however, continuously increases until and including the seventh time interval; the growing fraction of lower-temperature plasma shows the cooling of the flaring material. The smaller secondary peak in the hardness ratio lightcurve covered by the eighth spectrum makes $EM_3/EM_4$ temporarily decrease again.

\subsection{Abundance variations}

Even with two and three additional temperature components, residuals remain between 0.8--1.1~keV and and also at $\approx$2~keV, where many emission lines are located. This can be confirmed from the RGS spectra: Separating the RGS data into flare and quiescence, not only the continuum enhancement but also the flux increase in many emission lines in the wavelength range of 10--17~\AA\ becomes clearly visible. Since many of the lines located at these wavelengths originate from highly-ionized iron, this may however also be a temperature effect. In order to improve the fit and to disentangle temperature and abundance variations, we introduced single elemental abundances as additional free parameters to the sequence of fits. Particularly, we tested iron, neon, silicon and oxygen.

With the iron abundance as a free parameter, the maximum flare temperature is reduced to 32~MK; a strong enhancement in iron during the flare compensates part of the flux of the iron K complex, which serves as a major temperature indicator. The total emission measure only slightly decreases. However, the general behavior of the one-, two-, and three-temperature component fits remains unchanged, so the shape of both the flare temperature and the total emission measure lightcurves is preserved, as shown in Fig~\ref{EPIC_time}. With the iron abundance as a free parameter, the fit quality clearly improves. Table~\ref{epic_fit} lists the fit parameters for the optimum combination of two flare components for the first ten spectra and one component for the last spectrum. 

\begin{figure}
\resizebox{\hsize}{!}{\includegraphics{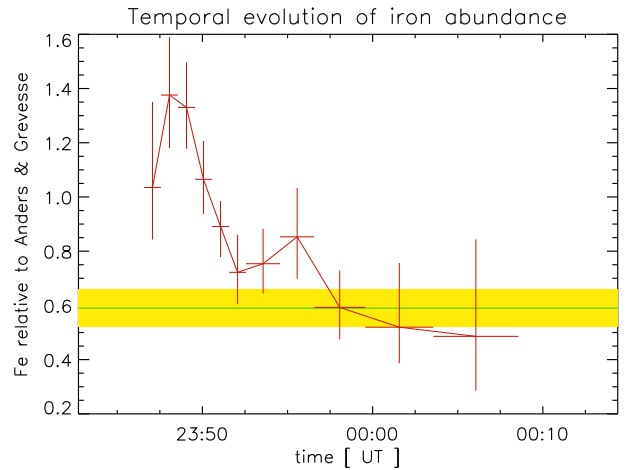}}
\caption[]{\label{Fe_time}Temporal evolution of the iron abundance (relative to \citet{Anders_Grevesse}). The quiescent value of 0.59$\pm$0.07 found by \citet{Fuhrmeister_CN_Leo} is shown for comparison.}
\end{figure}

Figure~\ref{Fe_time} shows the corresponding time series of the iron abundance relative to \citet{Anders_Grevesse} with the 1$\sigma$ confidence range of the quiescent abundance found by \citet{Fuhrmeister_CN_Leo} marked as a shaded area. The Fe abundance increases from a clearly subsolar level of 0.59$\pm$0.07 during quiescence by more than a factor of two to a maximum value of 1.38$\pm$0.20, when the flare peak is reached. It then decays even faster than the flare temperature; consistency with the quiescent value is reached again with the 6th spectrum. A re-increase at 23:55~UT in the eighth spectrum may not be significant, however, it coincides with the second rise of the hardness ratio lightcurve. By the end of the decay, the iron abundance has eventually dropped to the quiescent value again. A very similar behavior has been found by \citet{Algol_Flare} for the overall metallicity during a long-duration giant flare on Algol observed with \emph{BeppoSAX}.

Even with the Fe abundance as a free parameter, the second and the third spectrum cannot be fitted with a statistically acceptable value of $\chi^2_{red}$. Residuals remain mainly around 1~keV in the second and between 0.7 and 1.2~keV as well as at 2~kev in the third time interval, respectively. There are quite a few emission lines
at these energies, not only from iron but also from oxygen, neon and silicon. This indicates that further abundances may deviate from the quiescent values.
When the neon abundance is set free, it seems to be generally enhanced compared to the value of 1.20$\pm$0.16 from \citet{Fuhrmeister_CN_Leo}, the Ne abundance does however not give a clear pattern like the iron abundance in Fig.~\ref{Fe_time} but shows a large scatter, and the uncertainties are rather large. Most values are on the 1$\sigma$ level consistent with the old quiescent value. A free neon abundance therefore only marginally improves the fit, most of the residuals remain, with flare temperature and emission measure consistent stable at the initial values.
Silicon shows a similar behavior, with its abundance values scattering around the quiescent value of 0.78$\pm$0.33. Again, temperature and emission measure stay at the values obtained from the fitting sequence with all abundances fixed. The same procedure with oxygen tends to result in unreasonable fits with either the oxygen abundance or at least one of the temperature components diverging or approaching zero. 
We also combined two or more free elemental abundances, i.\,e. Fe and Ne, Fe and Si, as well as Fe, Ne and Si, but none of these combinations provided further improvement in the quality of the fit compared to Fe set free only. These fits even start to diverge, especially for the less well-exposed last two spectra with few degrees of freedom. Additionally, the residuals in the fits of the second and the third spectrum cannot be reduced significantly.

Iron therefore seems to be the only element, where a clear variation of the elemental abundance during the flare and compared to the quiescent state can be observed. In the common picture of stellar flares, the evidence of changes in the iron abundance indicates that the flare plasma, i.\,e. fresh material evaporated from photosphere and chromosphere, shows a different composition with a higher iron abundance compared to the ordinary coronal plasma. Since CN~Leo's quiescent corona shows the inverse FIP effect \citep{Fuhrmeister_CN_Leo}, this confirms the finding of \citet{Nordon_flares2} that IFIP-biased coronae tend to show a FIP bias during flares compared to their quiescent state. However, the Fe abundance seems to decrease much faster than temperature and emission measure of the flare plasma, and the iron abundance has returned to values consistent with the quiescent level when the X-ray emission of CN~Leo is still by far dominated by the flare. The time-resolved evolution of the iron abundance in the flare plasma suggests either the existence of some kind of iron-depleting mechanism in the flare plasma, which would be rather difficult to explain, or the possibility that different parts of the lower atmosphere with different abundance levels were affected during the early and later stages of the evaporation process.

Finally we would like to address the possibility that systematic errors in our fit models actually mimic an enhanced iron abundance.  At the very hot plasma temperatures as encountered during a flare, the iron abundance fit is affected by the flux of iron lines from lower ionization stages around 0.8--1.2~keV and  -- more importantly -- by the flux of the emission complex at 6.7~keV, mainly originating from iron in ionization stages from \ion{Fe}{xxiii} to \ion{Fe}{xxv}. An incorrect temperature model would lead to an incorrect prediction of the
emission both in the 0.8--1.2~keV energy range and in the 6.7~keV line complex, which would then have to be compensated by changing the iron abundance.  Whether this is the case or not is hard to tell, yet a detailed study undertaken by \citep{Algol_Flare} suggests that this is unlikely to happen.
Further, Fe K$\alpha$ fluorescence at 6.4~keV, excited in the neutral photospheric material by intense high-energy radiation $>$7.1~keV from the flare plasma, may also contribute. Fe K$\alpha$ fluorescence has been observed during strong stellar flares \citep[e.g.][]{Osten_II_Peg, Testa_fluorescence}.  The K$\alpha$ excitation depends on the photospheric iron abundance and on the inclination angle and height of the X-ray emitting flare plasma above the photosphere \citep{Testa_fluorescence, Ercolano_II_Peg, Drake_fluorescence}.  A compact loop structure as probably responsible for the short decay time of the flare on CN~Leo is expected to produce rather strong fluorescence,
yet no excess emission at 6.4~keV is visible, especially during the second and third time intervals, where the spectral fits yield the largest Fe abundance values.  All eleven spectra are well consistent with models without any Fe K$\alpha$ fluorescence emission at 6.4~keV. A possible explanation for this finding is an unfavorable viewing angle.
At any rate, the expected Fe K$\alpha$ line features are relatively weak, they can be spectrally separated from the 6.7~keV complex and should therefore not lead to incorrect Fe abundance estimates. 
Deviations from ionization equilibrium may also affect the strength of the 6.7~keV complex; \citet{Reale_nonequilibrium} have shown that the effect is significant for flares with heat pulse durations less than a minute. However, it is difficult to quantify this effect for the CN~Leo flare, also such deviations from ionization equilibrium are strongest at flare onset, while ionization equilibrium is established rather quickly with time in a high density environment as encountered with the flaring plasma in CN~Leo, so non-equilibrium effects appear to be an unlikely cause for the observed changes in iron abundance. Thus in summary, no clear systematic effects are known that would mimic an enhanced iron abundance during the
flare evolution.

\subsection{Densities}
\label{dens}
The RGS spectra allow to investigate the electron densities of the coronal plasma from the density-sensitive ratio of the forbidden and intercombination lines of helium-like triplets. With this method, \citet{Guedel_Proxima_Cen_1} found the density of the coronal plasma as traced by the \ion{O}{vii} and \ion{Ne}{ix} triplets to vary by more than an order of magnitude during a giant flare on Proxima Centauri, which is the only unambiguous measurement of increased coronal densities during a stellar flare. While the two flares were comparable in maximum countrate, the duration of the CN~Leo flare was much shorter, so that, in contrast to the Proxima Centauri flare and to the CN~Leo EPIC data, the CN~Leo RGS data can only be separated into quiescent state and flare; the RGS signal is too low for further sub-division. 

\begin{figure}
\resizebox{\hsize}{!}{\includegraphics{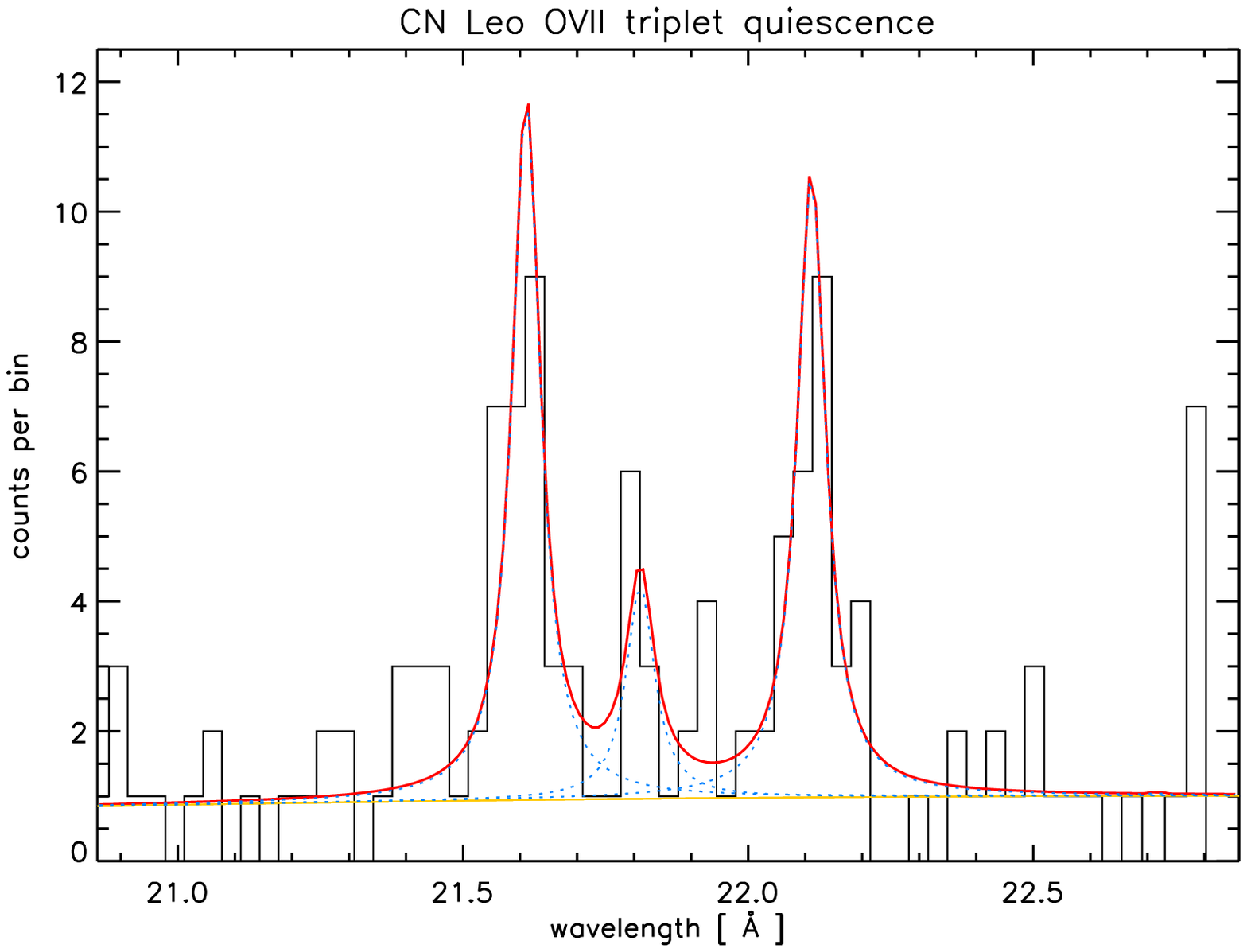}}
\resizebox{\hsize}{!}{\includegraphics{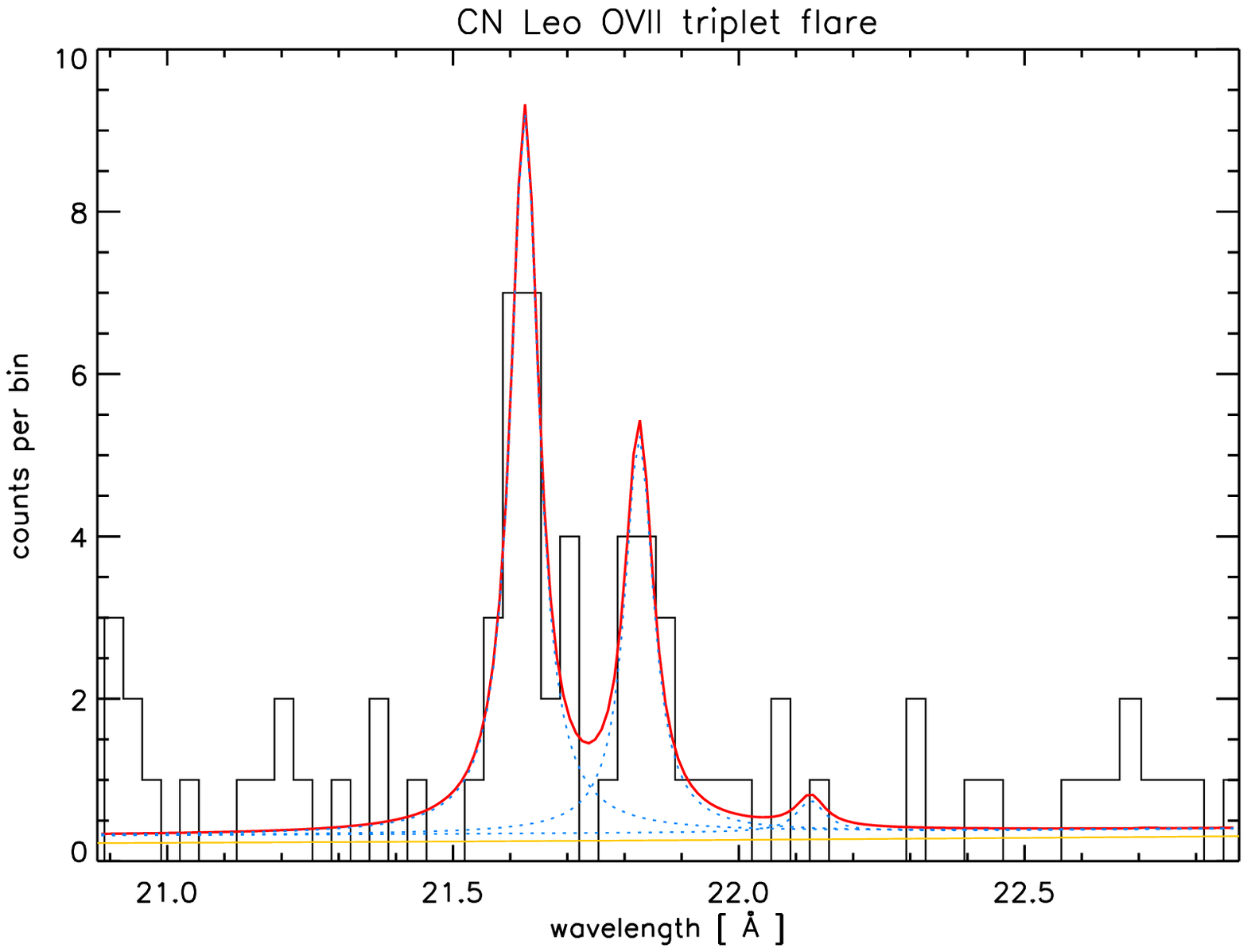}}
\caption[]{\label{OVII}The \ion{O}{vii} triplet in the RGS spectra of CN~Leo during quiescent (top) and flaring (bottom) state. The flare spectrum covers the same time interval as the eleven EPIC spectra, the quiescent data include the time intervals before and after the giant flare, with two further flares excluded.}
\end{figure}

\begin{table}
\begin{center}
\caption{\label{densities}Measured line counts, $f/i$ ratios and deduced coronal densities from the \ion{O}{vii} triplet in the RGS spectra of quiescence and giant flare in Fig.~\ref{OVII}.}
\begin{tabular}{lrrr}
\hline\hline
Line&Wavelength [\AA]& Quiescence& Flare\\
\hline
\ion{O}{vii} $r$	&21.60	&30.2$\pm$6.8	&25.4$\pm$5.9\\
\ion{O}{vii} $i$	&21.81 	&9.4$\pm$4.5	&14.0$\pm$4.7\\
\ion{O}{vii} $f$	&22.10	&27.1$\pm$6.3	&1.2$\pm$2.0\\	
\hline
\multicolumn{2}{l}{$f/i$}	&2.88$\pm$1.53	&0.09$^{+0.15}_{-0.09}$\\
\multicolumn{2}{l}{log $n_e$}	&$<$10.77	&$>$11.70\\
\hline
\end{tabular}
\end{center}
\end{table}

In Fig.~\ref{OVII} we show the RGS1 spectrum centered on the \ion{O}{vii} triplet during the 20~ks of accumulated quiescence and during the flare together with the best fits to the three triplet lines $r$ (resonance), $i$ (intercombination), and $f$ (forbidden) provided by the CORA program. The measured line counts, deduced $f/i$ ratios and electron densities are listed in Table~\ref{densities}. In order to convert the measured $f/i$~ratios to densities, we used the relation 
\begin{equation}
\frac{f}{i} = \frac{R_{0}}{1+n_{e}/N_{c}}
\end{equation}
with the critical density $N_{c}$ and the low-density limit $R_{0}$, where we adopted values of $3.1 \cdot 10^{10}$ cm$^{-3}$ and 3.95 from \citet{Pradhan_Shull}. During quiescence, the $f/i$ ratio is well consistent with the low-density limit, which is in good agreement with the values found for CN~Leo in May 2004 by \citet{Fuhrmeister_CN_Leo}. In the flare spectrum, the forbidden line has almost vanished, while the intercombination line is stronger. The electron density therefore clearly deviates from the quiescent value of $n_{\mathrm{e}}=1.2^{+4.9}_{-1.2} \cdot 10^{10}$~cm$^{-3}$ during the flare, and is definitely incompatible with the low-density limit. The average density during the flare has a nominal value of $n_{\mathrm{e}}=1.4 \cdot 10^{12}$~cm$^{-3}$. Taking into account the 1$\sigma$ error on the line fluxes, a lower limit of $n_{\mathrm{e}}>4.9 \cdot 10^{11}$~cm$^{-3}$, i.\,e. at least an order of magnitude higher than the quiescent value, can be derived. 

$f/i$ ratio and electron density obtained from the RGS flare spectrum represent mean values for the whole flare. In order to give a rough estimate of the peak flare density, one can derive a modifying scaling factor of $\approx$1.2 based on the square root of the maximum and mean total emission measure values from the EPIC data and assuming a constant plasma volume during the flare. The flux in the forbidden and intercombination lines measured during the whole flare is therefore dominated by the high countrates at the flare peak, and with a lower limit of $n_{\mathrm{e}}>5.9 \cdot 10^{11}$~cm$^{-3}$ the maximum density is only slightly larger than the average value.

With a peak formation temperature of $\approx$2~MK, the \ion{O}{vii} triplet traces only the cooler component of the coronal and flare plasma, and the density sensitivity of its  $f/i$ ratio is optimal between $10^{10}$ and $10^{11}$~cm$^{-3}$. As the \ion{O}{vii} triplet suggests densities probably even higher than $10^{12}$~cm$^{-3}$, it would be desirable to probe higher densities and the higher-temperature plasma with the helium-like triplets of heavier ions. Unfortunately, the flux in the RGS2 spectrum in the \ion{Ne}{ix} triplet, where the density-sensitive range has shifted to $10^{11}$--$10^{12}$~cm$^{-3}$ and the peak formation temperature has increased to $\approx$4~MK, is too low and its contamination with highly-ionized iron during the flare is too strong to give any conclusive result. \ion{Mg}{xi} and \ion{Si}{xiii} cover even higher densities up to $10^{14}$~cm$^{-3}$ and temperatures up to 10~MK, but in the two RGS spectra, the resonance lines of both triplets are only barely visible. A density survey of the flare at different plasma temperatures is therefore not possible. 

\subsection{The optical coronal \ion{Fe}{xiii} line}
With the RGS, the wavelength range $\approx$6--37~\AA\ can be observed; including individual emission lines with peak formation temperatures from $\approx$1.5--2~MK in the low-temperature range (\ion{C}{vi}, \ion{O}{vii}) up to 15~MK (\ion{Si}{xiv}). The optical coronal \ion{Fe}{xiii} line at 3388~\AA\ included in the blue-arm UVES spectra provides additional time-resolved information on the lowest coronal temperatures (peak formation temperature $\approx$1.5~MK).

Compared to the previous observations in December 2005, the \ion{Fe}{xiii} line is rather weak in all three observations in May 2006. The measured line fluxes have large uncertainties, sometimes the line is not even clearly detected. The measured FWHM is also smaller than previously observed. Unfortunately, no \ion{Fe}{xiii} line data is available for comparison from the observation in May 2004, where CN~Leo's quiescent X-ray emission was at a lower level, similar to the quiescent phase outside the large flare. However , it is reasonable to assume that the overall reduction in emission measure, which affects the lower temperatures where \ion{Fe}{xiii} is formed as well, is responsible for the lower line fluxes.

The most significant detections are obtained after the giant flare and after a second larger X-ray flare. A third larger X-ray flare is not covered by the optical observations anymore. The \ion{Fe}{xiii} line is not visible in the pre-flare spectrum and in the two flare spectra; but during the flare, a chromospheric \ion{Co}{i} line emerges at the same wavelength position. 

Even during quiescence, the bulk of iron in the coronal plasma should exist in the form of neon-like \ion{Fe}{xvii}, with only minor contributions of the cooler \ion{Fe}{xiii}. With the onset of the flare, the temperature of the X-ray emitting plasma shifts to even higher values, and therefore the ionization balance should be dominated by far by higher ionization stages. Even with an increased iron abundance, \ion{Fe}{xiii} can be expected to be depleted until the end of the decay, when the plasma has cooled down again. The excess emission measure then allows the detection of the \ion{Fe}{xiii}, which is consistent with the observations.

\section{Loop modeling}
Direct imaging demonstrates the great complexity of flaring active regions on the Sun. In order to obtain a consistent model describing the physical properties and evolution of the involved structures, a stellar flare can be reduced to a sequence of processes taking place in a simplified geometry. Such a generalized description starts with an initial heat pulse in a single coronal loop and efficient conduction until the whole loop has reached the maximum temperature. Lower atmospheric layers are strongly heated and chromospheric evaporation fills the loop with denser material. Conductive cooling sets in while the filling of the loop may still go on. Radiative cooling dominates as soon as the maximum density is reached, and loop depletion is initiated. Cooling and loop depletion may be decelerated by potential residual heating. The transitions between these processes can be described in terms of equilibrium conditions, but their possible overlap and interplay complicate the modeling. However, over the last decades, model approaches for flares have evolved from scaling laws valid only for (quasi)static conditions or simple analytical estimates to complex hydrodynamic loop models taking into account various initial conditions. These models have been successfully applied to a wide variety of solar and stellar flares.

\subsection{The decay phase}
\label{decay}
A basic finding for solar and stellar flares is that the decay time of a flare scales with the length of the flaring loop. In order to quantify this effect, \citet{Serio_flaremodel} developed an analytic approximation relating the thermodynamic decay time $\tau_{th}$, the loop half length $L$ and the loop top maximum temperature $T_0$ of a flaring loop, based on the underlying set of hydrodynamic equations and assuming semicircular loops with constant cross section cooling from equilibrium conditions and uniformly heated by an initial heat pulse without any further heating during the decay.
This approach has been refined by \citet{Reale_pre_loops} to account for loops comparable or larger than the pressure scale height $h$ at the loop top
\begin{equation}
h = \frac{k_B T_0}{\mu g}
\end{equation}
where $k_B$ is the Boltzmann constant, $\mu$ is the effective mass per particle and $g$ the surface gravity of the star. The pressure scale height for the giant flare is greater than 100\,000~km. 

In their models, \citet{Reale_pre_loops} also consider the effect of a gradually decaying heating function, and finally, for loops smaller than the pressure scale height, \citet{Reale_loops}  derived and tested an empirical expression to determine the flaring loop length including the effect of sustained heating during the decay that uses the slope $\zeta$ of the flare decay in the density-temperature plane to evaluate the amount of sustained heating during the decay:
\begin{equation}
\label{Eq_decay}
L = \frac{\tau_{LC} \sqrt{T_{\mathrm{0}}}}{\alpha \, F(\zeta)} \qquad \mathrm{or} \qquad L_9 = \frac{\tau_{LC} \sqrt{T_{\mathrm{0},7}}}{120 \, F(\zeta)}
\end{equation}
where $\alpha = 3.7 \cdot 10^{-4}$~cm$^{-1}$~s~K$^{1/2}$.

The term $F(\zeta)$ describes the ratio of observed and thermodynamic decay time $\tau_{LC}/\tau_{th}$ as a function of the ratio of temperature and density decay $\zeta$. The higher $F(\zeta)$, the larger is the amount of prolonged heating during the decay, while $F(\zeta) \approx 1$ would indicate the absence of additional heating. $F(\zeta)$ is monotonically decreasing and can be approximated with analytical functions; here we use the hyperbolic form 
\begin{equation}
F(\zeta) = \frac{\tau_{LC}}{\tau_{th}} = \frac{c_a}{\zeta-\zeta_a}+q_a 
\end{equation}
The coefficients $c_a$, $\zeta_a$, and $q_a$ depend on the energy response of the instrument used. For XMM/EPIC, the values are $c_a = 0.51$, $\zeta_a = 0.35$, and $q_a = 1.36$ according to \citet{Reale_diagnostics}. Additionally, the observed maximum of the flare temperature $T_{\mathrm{0,obs}}$, which is an averaged value over the whole loop, must be corrected with
\begin{equation}
\label{Eq_detector}
T_{\mathrm{0}} = \xi\ T^{\eta}_{\mathrm{0,obs}} \qquad \mathrm{or} \qquad T_{\mathrm{0},7} = \xi \cdot 10^{-7} \cdot T^{\eta}_{\mathrm{0,obs}}
\end{equation}
in order to obtain the true $T_{\mathrm{0}}$ independently of the detector. The values of the coefficients are $\xi = 0.13$ and $\eta = 1.16$ for XMM/EPIC \citep{Reale_diagnostics}. With the maximum flare temperature found in the second time interval, we determined $T_{\mathrm{0},7}$= 5.6$\pm$0.4.

\begin{figure}
\resizebox{\hsize}{!}{\includegraphics{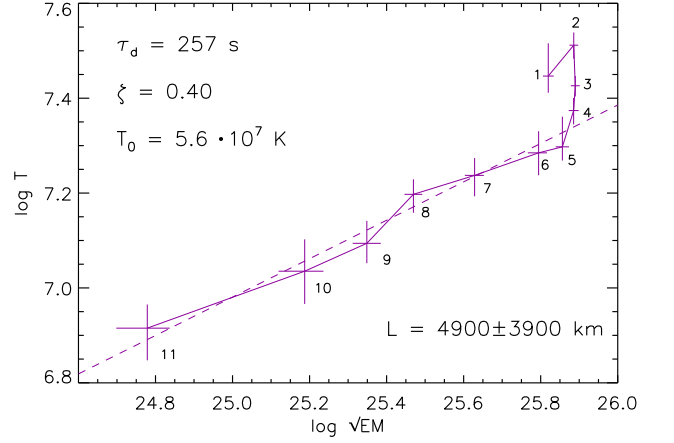}}
\caption[]{\label{EM_T}Flare evolution in the density-temperature plane.}
\end{figure}

Figure~\ref{EM_T} shows the evolution of the giant flare in the density-temperature plane, where we use $\sqrt{EM}$ as a proxy for the density. We fitted the slope $\zeta$ of the decay, i.\,e. with the data from first three spectra excluded, to $\zeta = 0.40 \pm 0.05$. This value is rather small, and on a 1$\sigma$ level consistent with the minimum value of 0.35, which means that the limit of validity for this method is reached. For very small $\zeta$, the corrective term $F(\zeta)$ that accounts for the presence of heating during the decay becomes very high and also rather uncertain, as the hyperbola approaches infinity. For $\zeta = 0.40$, we obtain a formal value of $F(\zeta) = \tau_{LC}/\tau_{th} = 10.7$; the thermodynamic decay time would therefore lie in the range of $\approx$~25 seconds, which is extremely short and hints to very effective cooling. The size of the involved loop structure is thus very small, the resulting loop half length is $L = 4900 \pm 3900$~km. The uncertainty in $L$ is dominated by far by the uncertainty in $\zeta$ in this flat part of the hyperbola of $F(\zeta)$. Effectively, we can therefore only give an upper limit for $L$, which can be set to $\approx$~9000~km using the 1$\sigma$ error.

The original approach of \citet{Reale_loops} is based on hydrostatic and energy equilibrium at the beginning of the decay. This would result in a simultaneous onset of the decay of temperature and density, which is apparently not the case for the giant flare on CN~Leo. In order to account for this, the flare temperature at the time of maximum density $T_{\mathrm{M}}$ should be used instead of $T_{\mathrm{0}}$ according to \citet{Reale_diagnostics}, otherwise the loop length will be overestimated. However, considering that the dependency of $L$ on the square root of $T_{\mathrm{0}}$ is comparably weak, we keep at the original value for the upper limit. Additionally, the exact point in time of maximum density is difficult to assess from flat top of the emission measure lightcurve, while the temperature already decreases significantly.

\subsection{Flare rise and peak}
Following the approach of \citet{Reale_diagnostics}, the duration of the flare rise and a possible delay between the decay of temperature and density can also be used to characterize the dimensions of a flaring loop. Both time scales can be traced back to  equilibrium states of heat input and cooling. The occurrence of such a delay, as it is also observed for the giant flare on CN~Leo, indicates an initial heat pulse too short to bring the flare plasma to equilibrium conditions, again pointing toward short time scales and a compact structure. 

The maximum density is reached when conductive and radiative cooling equal. This gives 
\begin{equation}
\label{Eq_rise}
L_9 = 3 \psi^2 \sqrt{T_{\mathrm{0,7}}}\ t_{\mathrm{M,3}}
\end{equation}
as an estimate for the loop length from time $t_{\mathrm{M}}$ at which the maximum density occurs and the ratio $\psi$ of the maximum temperature $T_{\mathrm{0}}$ and the temperature at maximum density $T_{\mathrm{M}}$:
\begin{equation}
\psi = \frac{T_{\mathrm{0}}}{T_{\mathrm{M}}}
\end{equation}
From the time $\Delta t_{\mathrm{0-M}} = t_{\mathrm{M}} - t_{\mathrm{0}}$ between the start of the temperature decay and the density maximum, i.\,e. the start of the density decay, one obtains
\begin{equation}
\label{Eq_peak}
L_9 = 2.5 \frac{\psi^2}{\ln \psi} \sqrt{T_{\mathrm{0,7}}}\ \Delta t_{\mathrm{0-M,3}}
\end{equation}
adopting the conductive cooling time.

These two independent approaches are particularly useful, if only the earlier phases of the flare are observed, i.\,e. Eq.~\ref{Eq_rise} can be applied if the complete rise phase of the flare is available until the density maximum is reached, while bigger parts of the decay are not required. Eq.~\ref{Eq_peak} can be used if additionally the quiescent phase before the flare and the flare onset itself are missing. When the flare is completely covered by the observations, the two methods yield independent estimates of the loop half length which can be cross-checked. 

Both Eqns.~\ref{Eq_rise} and \ref{Eq_peak} depend strongly on $\psi$ and also on the two onset times of the temperature and density decay, which have to be determined as accurately as possible. Using the time intervals from Sect.~\ref{spectra}, the temperature maximum occurs during the second spectrum, which would result in $t_{\mathrm{0,spec}} = 90 \pm 30$~s, i.\,e. a rather large uncertainty due to the length of the time interval compared with the rise time itself. It is, however, not reasonable to divide the sequence any further. The hardness ratio lightcurve from Fig.~\ref{hr} provides a better time resolution, assuming that especially during the early phases of the flare, the hardness ratio represents a good temperature indicator. The hardness ratio lightcurve undergoes major changes during the first time interval; it peaks already at $t_{\mathrm{0,HR}} = 47 \pm 5$~s, earlier than the spectral fit sequence, that provides only the average temperature of each interval. We would therefore prefer to use the hardness ratio as a temperature tracer, but subsequently, the true maximum temperature $T_{\mathrm{0}}$ will be larger than the 56~MK derived from the flare temperature of the second time interval, and it is difficult to give an accurate estimate. The difference between the two possibilities to choose for $t_{\mathrm{0}}$ will at any rate strongly affect the results of the loop length estimate.

It is even more difficult to define the maximum of the density. The emission measure, which we have used as the density indicator in Sect.~\ref{decay}, peaks in the third time interval. However, due to the flattened shape of the emission measure lightcurve, the uncertainty in $t_{\mathrm{M}}$ is probably even larger than the corresponding value of $150 \pm 30$~s. Consistent results are obtained when using a smoothed version of the countrate lightcurve, which would nevertheless result in an underestimate of  $t_{\mathrm{M}}$ according to \citet{Reale_diagnostics}. $T_{\mathrm{M}}$ is therefore determined from the flare temperature in the third time interval. 

Using the numbers obtained from the analysis of the time intervals only, we obtain $L = 25\,000 \pm 18\,400$~km from Eq.~\ref{Eq_peak}. When sticking to $T_{\mathrm{0}}$ from the second time interval but in combination with $t_{\mathrm{0}}$ from the hardness ratio lightcurve, $L$ increases to $43\,000 \pm 15\,300$~km, and a higher value for $T_{\mathrm{0}}$ would even result in a further increase, with the corresponding value being unknown.
Eq.~\ref{Eq_rise} gives $L = 16\,700 \pm 4900$~km, closer to the value from the upper limit from the analysis of the decay. Considering the large error bars, the results from Eqns.~\ref{Eq_decay}, \ref{Eq_peak}, and \ref{Eq_rise} are consistent at least on the 2$\sigma$ level. The overall uncertainties are dominated by the uncertainties of the two time scales $t_{\mathrm{0}}$ and $t_{\mathrm{M}}$: For the results obtained from the flare rise from Eq.~\ref{Eq_rise}, the uncertainty in $t_{\mathrm{M}}$ makes up 53\% of the uncertainty in $L$, while $T_{\mathrm{0}}$ and $T_{\mathrm{M}}$ share 33\% and 13\% respectively. When the time intervals are used to determine $t_{\mathrm{0}}$ in Eq.~\ref{Eq_peak}, $\Delta t_{\mathrm{0-M}}$ accounts even for 91\% of the uncertainty in $L$. Using $t_{\mathrm{0}}$ from the hardness ratio lightcurve, this reduces to 63\%. As Eq.~\ref{Eq_peak} yields indeterminate results, Eq.~\ref{Eq_rise} should be preferred.

\section{Discussion}
\subsection{Implications of the modeling}
\label{volumes}
The giant flare on CN~Leo has a very short exponential decay time of only a few minutes compared to other flares with similar amplitudes in X-rays, which typically last for hours. This holds even if one considers only flares on active M dwarfs \citep[e.\,g.][]{Favata_EV_Lac, Guedel_Proxima_Cen_1}. Such flares mostly originate from large loop structures with dimensions of at least a significant fraction of the size of the star itself. The short duration of the CN~Leo flare on the other hand points at short cooling times and therefore a rather compact involved structure. Respective estimates assuming a single flaring loop have confirmed this picture, the resulting loop half length is of the order of about a tenth of the stellar radius or even smaller.

We can assess the emitting plasma volume $V$ via 
\begin{equation}
V = \frac{EM}{n_e^2}
\end{equation}
and obtain the area $A$ of the loop footpoint on the stellar surface
\begin{equation}
A = \frac{V}{2L}
\end{equation}
approximating the loop with an unbent tube with a circular base area and constant cross-section over the loop length. In order to compute $V$ and $A$, we can use the electron density derived from the helium-like \ion{O}{vii} triplet of the flare spectrum in Sect.~\ref{dens}, and the total emission measure averaged over the eleven time intervals from Sect.~\ref{spectra}, $EM = 1.8 \cdot 10^{51}$~cm$^{-3}$, and obtain $V < 7.4 \cdot 10^{27}$~cm$^{3}$ from the lower limit of the density, or  $V = 9.1 \cdot 10^{26}$~cm$^{3}$ using the nominal value. This results in values of 1.3--$80 \cdot 10^{17}$~cm$^{3}$ for $A$, depending on the value used for $L$. Larger loop lengths, as they are provided by the data from the flare rise and peak, give the smallest footpoint area. A good reference point is $A \approx 1 \cdot 10^{18}$~cm$^{2}$, which would correspond to a short loop with a length close to the nominal value derived from the decay (i.\,e. $\approx$4500~km) at the nominal electron density value of $n_{\mathrm{e}}=1.4 \cdot 10^{12}$~cm$^{-3}$, or a longer loop with $L \approx$35\,000~km at a lower density of $5 \cdot 10^{11}$~cm$^{-3}$.

Another possibility to estimate $n_e$ independently from the emission measure is to make use of the scaling laws of \citet{RTV} 
\begin{equation}
T_0 = 1400 \sqrt[3]{pL}
\end{equation}
which assume equilibrium conditions and the plasma equation of state $p=2n_ek_BT$:
\begin{equation}
n_{\mathrm{0}} = 1.3 \cdot 10^{6} \frac{T^2_{\mathrm{0}}}{L} \qquad \mathrm{or} \qquad n_{\mathrm{0,10}} = 13 \frac{T^2_{\mathrm{0,7}}}{L_9} 
\end{equation}
This results in density values between $n_{\mathrm{e}}=3.9 \cdot 10^{11}$~cm$^{-3}$ and $n_{\mathrm{e}}=2.8 \cdot 10^{12}$~cm$^{-3}$ for the largest and smallest possible loop lengths respectively, which matches the maximum density estimate from Sect.~\ref{dens} quite well. The maximum density $n_M$ should be somewhat higher than $n_0$. However, due to the large uncertainties in $L$, the derived densities cannot be fixed to an order of magnitude. Similarly, footpoint areas between $A = 7.7 \cdot 10^{17}$~cm$^{3}$ and $A = 3.2 \cdot 10^{18}$~cm$^{3}$ are obtained, but again with large uncertainties on individual values.
In addition, it is unclear to which level this particular flare follows the established scaling laws at all, as the initial assumptions like equilibrium or constant flaring volume are not necessarily fulfilled.

In order to model the short initial burst, \citet{CN_Leo_onset} assumed a single loop with a half length of 20\,000~km. Along the loop, they obtain densities of 1--$10 \cdot 10^{11}$~cm$^{-3}$ after it is filled with evaporated material. The temperature in the loop model reaches maximum values around $4 \cdot 10^{7}$~K, both numbers are well consistent with what we obtain also for the major flare event.

\citet{Aschwanden_laws} compiled statistical correlations between peak temperature $T_0$, peak emission measure $EM$, X-ray luminosity $L_X$, total X-ray-radiated energy $E_X$, and flare duration $\tau_f$ for a large sample of observed stellar flares. They find the following relations as a function of the maximum temperature $T_0$, 
\begin{equation}
\label{emission}
EM(T_0) = 10^{50.8} \cdot T_{0,7}^{4.5\pm0.4}
\end{equation}
\begin{equation}
\label{duration}
\tau_f(T_0) = 10^{2.5} \cdot T_{0,7}^{1.8\pm0.2}
\end{equation}
\begin{equation}
\label{luminosity}
L_X(T_0) = 10^{27.8} \cdot T_{0,7}^{4.7\pm0.4}
\end{equation}
\begin{equation}
\label{energy}
E_X(T_0) = 10^{30.7} \cdot T_{0,7}^{6.1\pm0.5}
\end{equation}
however, the correlation between $\tau_f$ and $(T_0)$ in Eq.~\ref{duration} is only marginally significant due to large scatter in the data. Our measurements from giant flare on CN~Leo clearly deviate from the relation between peak temperature and emission measure, the observed maximum emission measure is more than an order of magnitude lower than the value provided by Eq.~\ref{emission}. While the densities we observe seem to be common for such a huge flare, the flaring structure is rather compact, i.\,e. the flaring volume is very small, which apparently causes the low emission measure. Eq.~\ref{duration} predicts a flare duration of about 45 minutes, more than two times longer than observed. However, as the correlation between $\tau_f$ and $T_0$ is not that strong, this matches the data regime from the flare sample. As a consequence of the deviation of the emission measure, Eq.~\ref{luminosity} overestimates the X-ray luminosity at the flare peak also by an order of magnitude. The total radiative loss in X-rays from Eq.~\ref{energy} is too high even by two orders of magnitude as a result of both the overestimated emission measure and flare duration.

\subsection{Single loop or arcade?}
The Sun shows us that the model of an accumulation of isolated single loops oversimplifies a stellar corona. Instead, loops of different sizes often group to complex structures of arcades, especially in the active regions where flares originate. Such loop structures are non-static; individual loops resize, their footpoints move, and they realign in magnetic reconnection events.
Unfortunately, we cannot even resolve individual active regions directly on stars other than the Sun. Doppler imaging is a common indirect method to reconstruct photospheric spots or prominences in the chromosphere of fast rotators spectroscopically in the optical \citep[see e.\,g][]{Strassmeier_review}. Lightcurve-based eclipse mapping techniques provide an additional approach, as successfully demonstrated by \citet{Wolter_CoRoT2}. However, the solutions of such image reconstruction methods are usually not unique, becoming more and more ambiguous the better the targeted spatial resolution is. Additionally, such observations are very rare in the X-ray regime, cf. \citet{Guedel_YY_Gem, Guedel_alpha_CrB}, and \citet{Algol_RGS}.

There are however other clear indications for the substructuring of active regions in stellar coronae. Flare lightcurves often deviate from the simple scheme of rise and exponential decay, and instead show multiple peaks or variations in the decay time, like a double exponential decay \citep[e.\,g.][]{Wargelin_Ross_154}. The lightcurve of the giant flare on CN~Leo shows several irregularities like the initial outburst and the flat peak. Its decay deviates from an exponential shape, and there is the small bump in the lightcurve which is apparently related to a second peak in the hardness ratio lightcurve. This event obviously interrupts the otherwise continuous cooling of the flare plasma. Such features suggest that multiple components of the active region from which the whole flare event originates are successively involved, each of them with its own properties like shape and size, temperature and density, and hence cooling time. 

Further evidence for flare substructure can be obtained from the spectral analysis of the flaring plasma. A strong correction $F(\zeta)$, i.e. when heating dominates the decay of the flare, points to an arcade of loops involved, as observed for two-ribbon flares on the Sun; this was first pointed out by \citet{Reale_flaremodeling}. For the giant flare, $F(\zeta)$ is very high. We therefore interpret the whole event in terms of a flaring arcade. In this scenario, the flat top of the lightcurve can be explained by the superposition of several loops flaring consecutively. Later during the global decay, further loops not affected so far became involved and caused the small bump. This flare series was triggered by the short initial outburst, which could be localized in an almost isolated single loop \citep{CN_Leo_onset}. 
If we assume that the flat peak of the main flare is composed of a sequence of unresolved individual peaks with similar luminosity, the true values of $t_{\mathrm{M}}$ and in particular $\Delta t_{\mathrm{0-M}}$ would be lower while $\tau_{LC}$ will stay roughly the same, which would in turn reduce $L$ from Eqns.~\ref{Eq_rise} and \ref{Eq_peak}, i.e. bring the results from the analysis of rise and peak closer to the upper limit obtained from the decay.

A theoretical model framework for solar two-ribbon flares has been developed by \citet{Kopp_Poletto} and -- with restrictions -- successfully applied also to stellar flares, see e.g. \citet{Poletto_modelflares} for the modeling of two long-duration flares on the active M dwarfs EQ~Peg and Proxima Centauri observed with \emph{Einstein} and \emph{EXOSAT}. The single-loop model approach of \citet{Reale_loops} can be extended to multi-loop structures if different phases of the flare are analyzed separately. \citet{Reale_Proxima_Cen} investigated the evolution of the long-duration giant flare on Proxima Cen observed with \emph{XMM-Newton}. In their best-fit model, the X-ray emission is dominated by a single loop for the first part of the flare, while later on during the decay an additional arcade of loops with approximately the same length as the initial single loop emerges. This picture is supported by the analysis of spatially resolved observations of strong solar two-ribbon flares  \citep[e.g.][]{Aschwanden_Alexander}. For the giant flare on CN~Leo, the primary single loop was causing the 2-second initial outburst. Due to its short duration it is impossible to analyze this very first part of the flare in more detail in order to obtain flare temperature and loop length. We can, however, conclude that even if the whole flare event is treated as originating from an arcade-like structure, it would be very compact, probably even smaller in terms of loop length than a single loop. \citet{Reale_Proxima_Cen} found a loop length larger by 30\% if the flare on Proxima Cen is interpreted with Eqns.~\ref{Eq_decay}--\ref{Eq_detector} instead of with detailed hydrodynamic loop modeling. For a single loop, the flaring volume scales with $L$, while an arcade of loops implicates $V \propto L^3$. For solar flares, an empirical correlation $V \propto L^{2.4}$ is obtained \citep{Aschwanden_laws}, reflecting the observation that arcades are not uniformly filled but consist of individual loops. 

\subsection{The multiwavelength context}
Plasma temperature, density, and optical thickness of the stellar atmosphere change by orders of magnitude, when going from the photosphere via chromosphere and transition region up into the corona.
According to its spectral type, i.e. the effective temperature, the major contribution to the photospheric spectrum of an M~dwarf like CN~Leo is emitted the infrared, with a much smaller proportion observed in the optical. The chromosphere adds line emission, mainly in the UV and in the optical. Line formation at these wavelengths is a complex process. Photospheric absorption lines may turn into emission in the chromosphere, and the strength and shape of chromospheric lines is extremely sensitive to the atmospheric conditions. A sophisticated modeling of M dwarf chromospheres is challenging, and current models are still far from providing a complete description \citep[see e.g.][]{Fuhrmeister_Mdwarfmodels}. 
Nevertheless, as for each line the formation takes place under specific physical conditions which basically translate into certain heights above the the ``surface'', different lines probe the structure of the stellar atmosphere. Transition region and corona emit mainly in X-rays and in the EUV. Complemented by non-thermal processes like e.g. radio gyrosynchrotron emission, the different atmospheric layers can therefore be studied over a broad wavelength range.

Solar and stellar flares can affect all parts of the atmosphere, with the strongest impact observed on chromosphere and corona. However, as on M dwarfs the quiescent photospheric continuum is low in the optical, photospheric flare continuum emission, that accompanies strong events like the giant flare on CN~Leo discussed in our series of papers, is much more obvious compared to the Sun, where white light flares are rarely observed. When integrating over the full disk, they would go unnoticed in optical broad band measurements. In Paper~I, we analyzed the observed optical continuum emission during the CN~Leo flare. We found that the photospheric material was rapidly heated to temperatures probably exceeding 20\,000~K at the onset of the flare and rapidly cooling afterwards. The size of the affected region in the photosphere agrees very well with the area derived in Sect.~\ref{volumes}. Our chromospheric modeling in Paper~II confirmed the strong heating of the lower atmospheric layers, during the flare the chromosphere is shifted towards higher column masses, i.e. deep into the photosphere. In Paper~I we presented a detailed list of emission lines observed in the optical flare spectra; the included lines provide good indicators for the structure of chromosphere and transition region. However, as shown in Paper~II, a self-consistent model of the flare and especially of its early phases would require to consider also non-equilibrium conditions. Additionally, the flare plasma is non-static, line shifts and asymmetries are tracers of mass motions, i.e. chromospheric evaporation into the corona and later on material that has cooled down and falls back towards the surface. Unfortunately, the limited spectral resolution of the X-ray data does not allow to verify this for the hot plasma in the corona. 

Nevertheless, our multiwavelength study has revealed a very detailed picture of the flare. We were able to determine plasma temperatures, densities and abundances, apart from the latter not only in the corona but also in photosphere and chromosphere. We learned that the active region and its associated coronal loop structure where the flare originated were rather compact but probably also much more complex than a simple single loop with two footpoints. 
Multiwavelength observations can provide a comprehensive picture of stellar activity phenomena. \citet{Wolter_multiwavelength} have complemented their X-ray observations of the fast rotator Speedy~Mic with Doppler imaging information, which allowed to localize not only spots on the stellar surface but also prominences and a flare. As theoretical models become more and more complex, with both stellar atmosphere codes and hydrodynamic simulations approaching the non-static 3D domain, including magnetic fields or further physical conditions, only detailed observing data can provide complementary information. For stellar flares, this requires good spectral and temporal resolution over a wavelength range as broad as possible.

\section{Summary and conclusion}
We have analyzed simultaneous X-ray and optical data covering a giant flare on the active M dwarf CN~Leo with regard to the plasma properties in the corona and the geometry of the flaring structure. As it could be expected from other events of similar strength observed in X-rays, temperature and density are enhanced by more than an order of magnitude compared to the quiescent corona at the flare peak. The flare plasma, which can be considered to consist mostly of material evaporated from chromosphere or photosphere, shows a different composition. Despite its high amplitude, the flare was only of short duration, indicating a rather compact flaring structure, which is also comfirmed by simple loop modeling. Sustained heating during the decay as well as substructure in the lightcurve and hardness ratio however suggest an arcade-like structure much more complex than a single flaring loop. Broad multiwavelength coverage allowed us to characterize this exceptional event in great detail.

\begin{acknowledgements}
We would like to thank our referee, Fabio Reale, for his valueable suggestions that helped to improve the Paper.
C.L. and B.F acknowledge financial support by the DLR under 50OR0105.
\end{acknowledgements}

\bibliographystyle{aa}
\bibliography{../literature}

\end{document}